\documentclass[%
 aip,
 amsmath,amssymb,
 reprint,%
]{revtex4-1}

\usepackage{graphicx}
\usepackage{dcolumn}
\usepackage{bm}
\usepackage[utf8]{inputenc}
\usepackage[T1]{fontenc}
\usepackage{mathptmx}
\usepackage{etoolbox}

\usepackage{hyperref}
\hypersetup{
	colorlinks=true,       
	linkcolor=blue,          
	citecolor=blue,        
	filecolor=magenta,      
	urlcolor=blue
	}

\makeatletter
\def\@email#1#2{%
 \endgroup
 \patchcmd{\titleblock@produce}
  {\frontmatter@RRAPformat}
  {\frontmatter@RRAPformat{\produce@RRAP{*#1\href{mailto:#2}{#2}}}\frontmatter@RRAPformat}
  {}{}
}%
\makeatother

\preprint{AIP/123-QED}

\begin{document}

\preprint{AIP/123-QED}

\title[]{Quantum fluctuations spatial mode profiler} 

\author{Charris Gabaldon}{}
 \affiliation{Physics Department, William $\&$ Mary, Williamsburg, Virginia, USA}
\author{Pratik Barge} 
 \affiliation{Hearne Institute for Theoretical Physics and Department of Physics $\&$ Astronomy, Louisiana State University, Baton Rouge, Louisiana, USA}
\author{Savannah L. Cuozzo}
 \affiliation{Physics Department, William $\&$ Mary, Williamsburg, Virginia, USA}
\author{Irina Novikova}
 \affiliation{Physics Department, William $\&$ Mary, Williamsburg, Virginia, USA}
\author{Hwang Lee}
\affiliation{Hearne Institute for Theoretical Physics and Department of Physics $\&$ Astronomy, Louisiana State University, Baton Rouge, Louisiana, USA}
\author{Lior Cohen}
\affiliation{Department of Electrical, Computer and Energy Engineering, University of Colorado Boulder, Boulder, 80309, CO, USA}

\author{Eugeniy E. Mikhailov}
 \affiliation{Physics Department, William $\&$ Mary, Williamsburg, Virginia, USA}


\date{\today}

\begin{abstract}
The spatial mode is an essential component of an electromagnetic field description, yet it is challenging to characterize it for  optical fields with low average photon number, such as in a squeezed vacuum. We present a method for reconstruction of the spatial modes of such fields based on the homodyne measurements of their quadrature noise variance performed with a set of structured  masks. We show theoretically that under certain conditions we can recover individual spatial mode distributions by using the weighted sum of the basis masks, where weights are determined using measured variance values and phases. We apply this approach to analyze the spatial structure of a squeezed vacuum field with various amount of excess thermal noise generated in Rb vapor. 

\end{abstract}

\maketitle
\section{Introduction}
Transverse spatial distribution is an important element of the description of any classical or quantum electromagnetic field. For many applications it is essential to restrict light propagation within a single, well-defined spatial mode. However, the multimode nature of light can be desirable in fields such as optical information multiplexing~\cite{Gibson2004OEoams,qCom} or imaging~\cite{PhysRevLett.100.143601,multimodetwinGenovese2011,PhysRevA.105.023725}. In either case the ability to identify and characterize the spatial mode composition of an electromagnetic field becomes a helpful tool. Several solutions have been recently proposed for classical optical fields, in which specially designed dispersive elements help spatially separate various modes (often in Laguerre-Gauss or Hermit-Gauss basis) into uniquely positioned spots~\cite{Berkhout2010PRLoamsorting,ZhouOL18,FuOE18,fontaineNC2019}. The situation becomes significantly more challenging when the multimode optical field consists primarily of squeezed vacuum quantum fluctuations, since there is no accompanying strong classical field to tune to a selected mode. In this case identification of individual modes becomes akin looking for a black cat in a dark room. Traditional quantum noise detection requires a strong local oscillator (LO) to amplify weak fluctuations to the detectable level.
However, this method relies on perfect spatial overlap of the LO and the unknown quantum probe~\cite{Gerry_and_Knight_book,Bennink_Boyd_PhysRevA}, and thus requires \textit{a priori} knowledge of the quantum field's transverse distribution, or the perfect shape of the LO need to be found via the set of optimization measurements.
The situation becomes more complicated if the quantum noise is spatially multimode, such as a mixture of, e.g., squeezed vacuum and thermal light.
In some cases, e.g., if the squeezed modes are not overlapping, it is possible to obtain the information about their number, shapes and squeezing parameters by reducing the size of the LO mode~\cite{lettSci08,Embrey2015,MarinoPhysRevA.98.043853} or sampling nearby pixels correlations~\cite{MarinoPhysRevA.100.063828} as it was demonstrated for twin-beam squeezing. Multimode quantum field is a useful resource for quantum imaging, as the information about spatial transmission masks can be obtained by shaping the LO~\cite{MarinoEJPD2012,Clark2012,lettSci08} or analyzing noise correlation for each camera pixel~\cite{pratik2022,Cuozzo2022camera}. However, these measurements rely on the relative modification of the quantum probe noise, and may not be useful for the diagnostic of the original multimode probe itself. 
The Bloch-Messiah reduction~\cite{Bloch1962,braunstein2005squeezing, horoshko2019bloch} offers a promising method to extract information about squeezing modes of a multimode optical field. It was shown to recover the set of quantum eigenmodes of the frequency comb~\cite{Medeiros2014,Cai2017} and parametric amplifiers via the diagonalization of the measurement basis. However, this is a data intensive procedure - the required number of measured covariances is proportional to the square of the measurement basis elements.

Here we propose a protocol for characterizing and reconstructing the spatial profiles of single- and two mode quantum fluctuations with no prior information. While not as general as the Bloch-Messiah reconstruction, our method is significantly simpler since the required number of measurements scales linearly with the number of basis elements (spatial pixels). In particular, we reconstruct a transverse distribution of a single squeezed vacuum mode, and then expand the formalism to describe  the mixture of the squeezed and thermal modes. Our method is based on single pixel imaging techniques~\cite{Clemente_Single_Pixel_Holography_2013,Sidorenko2016,Gibson2020,Li2021,Sephton2023} adapted to the quantum domain and combining it with homodyne detection~\cite{Cuozzo2022spi}. Full wavefront information about the phase and amplitude in each point is extracted from the quantum quadrature variance measurements. In our experimental reconstruction we use a quadrature squeezed vacuum source based on the PSR nonlinearity in Rb atoms~\cite{matsko_vacuum_2002,ries_experimental_2003,mikhailov2008ol}, and trace the modification of the output quantum state from mostly single-mode squeezed vacuum to an admixture of squeezed vacuum and excess thermal noise as the temperature of Rb vapor increases~\cite{hsu_effect_2006,mikhailov2009jmo,mizhang2016pra,nicktheorypaper}. However, our method is general and can be adopted for wide range of squeezed light sources and wavelengths.

\begin{figure*}
    \centering
    \includegraphics[width=0.95\linewidth]
    {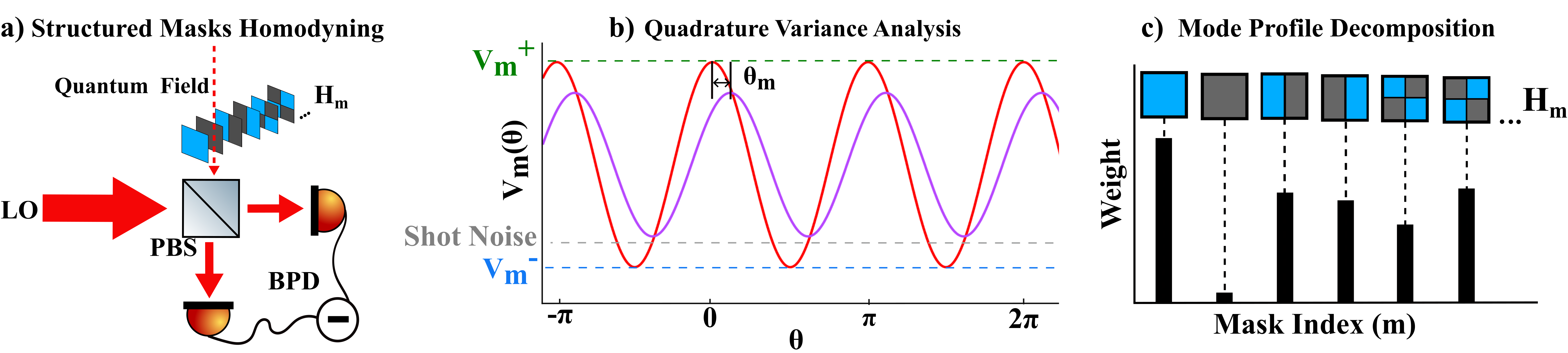}
    \caption{
    General concept of the noise profiler: (a) a full set of Hadamard masks ($H_m$) is applied to the quantum field before its variance is measured using a homodyne detector. 
    (b) The measured quadrature noise is recorded as a function of a LO phase, and values for the maximum and minimum noise variances $V^\pm_m$, as well as the shift in the quadrature phase $\theta_m$, as defined by Eq.~(\ref{eq:measured_variance_simplified_form}), are extracted for each mask $H_m$. (c) A complex weight value is calculated for each Hadamard mask, using the measured variance values, such that the weighted sum of $H_m$ recreates the shaped squeezed field, as defined by Eq.~(\ref{eq:shapedsqzfield}).
    }.
    \label{fig:sqz_detection}
\end{figure*}

The general idea of our quantum noise mode profiler is inspired by classical single-pixel imaging~\cite{Gibson2020} combined with the homodyne detection~\cite{Cuozzo2022spi}. The principle difference is that we detect the quadrature noise variance, rather than average light power, of the optical field  after a set of spatial transmission masks, as illustrated in Fig.~\ref{fig:sqz_detection}.(a). For each squeezed mode the mask changes its quantum noise by reducing their quadrature fluctuations as well as their squeezing angle. We trace these changes for each mask $H_m$ using the homodyne detector. By analyzing the variance as a function of the LO phase, we can find minimum and maximum variance $V^\pm_m$ as well as the relative phase $\theta_m$ [Fig.~\ref{fig:sqz_detection}(b,c)] and use them to calculate the weights to reconstruct the original signal spatial profile using the masks [Fig.~\ref{fig:sqz_detection}(d)]. Notably, the same procedure will work if the masks are placed on the LO, rather than the signal field. 


\section{Quadrature variance calculations for multimode quantum fluctuations }
In this section we analytically calculate the quadrature variance of a Gaussian optical quantum state with multiple spatial modes overlapped with a LO in a balanced homodyne measurement setup and validate the mode reconstruction method from the measured noise values.
To calculate the homodyne detection output in the multimode case, we need a powerful formalism that can efficiently handle the multimode complexity. Fortunately, we can model the signal quantum field as N-mode Gaussian states -- continuous variable (CV) states with Gaussian Wigner function~\cite{weedbrook2012gaussian}: 
\begin{equation}
    W(x) = \frac{\exp{\left[-\frac{1}{2}(\hat{x}-\left\langle\hat{x}\right\rangle)^T V^{-1}(\hat{x}-\left\langle\hat{x}\right\rangle)\right]}}{(2\pi)^N \sqrt{det V}}.
\end{equation}
These states are completely determined by their first two moments, the mean vector $\hat{x} = (\hat{q}_1,\hat{p}_1,...\hat{q}_N,\hat{p}_N)^T$ and covariance matrix
\begin{equation}
    V_{jk} = \frac{1}{2}\left\langle \left\{ (\hat{x}_j-\left\langle\hat{x}\right\rangle_j),(\hat{x}_k-\left\langle\hat{x}\right\rangle_k) \right\} \right\rangle,
    \label{eq:cov_evolve}
\end{equation}
where \{*,*\} denotes anticommutator, $\hat{q}_k = \frac{1}{\sqrt2}(\hat{a}^\dagger_k+\hat{a}_k)$ and $\hat{p}_k=\frac{1}{\sqrt2i}(\hat{a}^\dagger_k-\hat{a}_k)$ are the quadrature operators associated with the k${th}$ mode defined via standard creation $(\hat{a}^\dagger_k)$ and annihilation $(\hat{a}_k)$ operators. Diagonal elements of the covariance matrix represent the quadrature variance of the field modes.
%
For example, a signal field consisting of two squeezed vacuum modes with different spatial profiles is represented by a $4\times4$ covariance matrix  $V^{(in)} = \big(\begin{smallmatrix}v_1 & 0\\ 0 & v_2\end{smallmatrix}\big)$ with $k \in \{1,2\}$ and
\begin{equation}
\label{eq:gaussian_state_covariance_matrix}
v_k = 
\begin{pmatrix}
e^{r_k}\cos^2\phi_k+ e^{-r_k}\sin^2\phi_k& [e^{-r_k}-e^{r_k}]\cos\phi_k\sin\phi_k \\
[e^{-r_k}-e^{r_k}]\cos\phi_k\sin\phi_k & e^{r_k}\sin^2\phi_k+ e^{-r_k}\cos^2\phi_k,
\end{pmatrix}
\end{equation}
where $r_k$ and $\phi_k$ are the squeezing parameter and squeezing angle for each mode.

Next, we need to describe the transformation of the quantum state after the mask and predict the output of the homodyne detector. We can model these optical elements using two symplectic matrices. Matrix $B$  models a mask as a beam splitter with transmission $T$, and matrix $R$ represents the  single mode phase rotation $\theta$:
\begin{equation*}
B = 
\begin{pmatrix}
\sqrt{T}&  0&  \sqrt{1-T}& 0 \\
 0&  \sqrt{T}&  0& \sqrt{1-T} \\
 \sqrt{1-T}&  0&  -\sqrt{T}&  0\\
 0&  \sqrt{1-T}&  0& -\sqrt{T}
\end{pmatrix},
\end{equation*}
\begin{equation}
R = 
\begin{pmatrix}
\cos\theta&  \sin\theta&  0& 0 \\
 -\sin\theta&  \cos\theta&  0& 0 \\
 0&  0&  1&  0\\
 0&  0&  0& 1
\end{pmatrix}.
\end{equation} 

The first diagonal matrix element of the final covariance matrix~\cite{symplectictransform} provides the value of the the output noise quadrature at the output of the homodyne detection $V(\theta)$:
\begin{equation}
V(\theta)=V^{(out)}_{11}=R_{1j}B_{jk}V^{(in)}_{kl}B^{\dagger}_{lm}R^{\dagger}_{m1}=R_{1j}R_{1m}B_{jk}B_{ml}V^{(in)}_{kl}\,,
\end{equation}
where we connect the initial and final covariance matrices by applying the transformation with matrix multiplication. 
Taking into account the positions of the nonzero elements of the matrices involved, indices $j$ and $m$ can be only 1 and 2 and consequently indices $k$ and $l$ can be 1 and 3 or 2 and 4. This simplifies the matrix product to only 8 nonzero terms:
\begin{multline}
V(\theta)=\sum_{j,m,k,l} R_{1j}R_{1m} T \delta_{jk}\delta_{ml}V^{(in)}_{kl} \\ + \sum_{j,m,k,l}R_{1j}R_{1m}(1-T)\delta_{j+2,k}\delta_{m+2,l}V^{(in)}_{kl}\nonumber
\end{multline}




The beam splitter matrix $B$ accounts for the transformation of each of the squeezed modes by the mask. The transformation (matrix) coefficients are exactly the overlap between the input ($u_k(x,y)$) and output ($H(x,y)$) modes which is defined by the integral:
\begin{align}
{\cal O}_k = \int u_{k}(x,y) H(x,y) dx dy.
\end{align}
 In this case of two modes, $T=|{\cal O}_1|^2$ , $1-T=|{\cal O}_2|^2$, and the final expression for the two single mode squeezing signal variance after the mask becomes:
\begin{align}
\label{eq:noise2modes}
V(\theta)= 1 + \sum_{k = 1}^{N=2} |{\cal O}_{k}|^2 \left[e^{r_k}\cos^2(\theta-\phi_k) + e^{-r_k}\sin^2(\theta-\phi_k)-1\right].
\end{align}
This result can be easily generalized to $N$ modes, by applying $N$-way beam splitter transformation and absorbing the phases, induced by the transformation, into the squeezing angles. 

Next, we can further extend the treatment into the squeezed thermal states. The covariance matrix is changed by adding additional factors $2\bar n_{{\rm th},k}$ to all diagonal terms, where $\bar n_{{\rm th},k}$ is the average thermal photon number in the $k$th mode. Combining all this into Eq.~\ref{eq:noise2modes} we get:
\begin{multline}
\label{eq:noiseNmodes2}
V(\theta)= 1 + \sum_{k = 1}^{N} |{\cal O}_{k}|^2  \big[(e^{r_k}+2\bar n_{{\rm th},k})\cos^2(\theta-\phi_k) + \\ +
(e^{-r_k}+2\bar n_{{\rm th},k})\sin^2(\theta-\phi_k)-1\big].
\end{multline}
It is easy to see that the results in Eqs.\ref{eq:noise2modes} and \ref{eq:noiseNmodes2} can be written in the same general form:
\begin{align}
	\label{eq:detected_variance_multimode_1m}
	V(\theta) & = 1 + \sum_k |{\cal O}_{k}|^2 \left( V^+_k
	\cos^2(\tilde\theta_{k}) + V^-_k\sin^2(\tilde\theta_{k}) -1 \right), \\
\tilde{\theta}_{k} &= \theta - \phi_k - Arg{{\cal O}_{k}},
\end{align}
if we identify $V_k^+=e^{r_k}+2\bar n_{{\rm th},k}$ and $V_k^-=e^{-r_k}+2\bar n_{{\rm th},k}$.

\section{Unknown spatial mode reconstruction via quadrature noise measurements}

Now we are ready to discuss the reconstruction of the unknown signal mode profile by measuring its quadrature variance after a complete set of transmission masks $H_m(x,y)$. In general, the signal may consist of multiple spatial modes, each described by $u_k(x,y)$. For the purpose of this discussion, we will assume that the quantum fluctuations of each modes are defined by its maximum and minimum quadrature noise $V^\pm_k$ (normalized to the vacuum state
noise), and $\theta_k$ is the squeezing angle with respect to the local oscillator, that we assume to be a single-mode coherent field with the spatial distribution $u_{LO}(x,y)$. 


To gain information about the spatial profile of the input field
we modify the signal field by passing it through various mask $H_m(x,y)$ and measuring the corresponding 
quadrature variance $V_m(\theta)$:
\begin{align}
	\label{eq:detected_variance_multimode}
	V_m(\theta) & = 1 + \sum_k |{\cal O}_{km}|^2 \left( V^+_k
	\cos^2(\theta - \theta_k) + V^-_k\sin^2(\theta - \theta_k) -1 \right) ,
\end{align}
where
\begin{align}
	\label{eq:mask_overlap_integral}
	{\cal O}_{km} = \int_{A_d} u_{LO}^*(x,y) u_k(x,y) H_m(x,y) dx dy\,.
\end{align}
Note that such a mask can instead be introduced into the LO path as it
would not change above overlap parameter definition except the mask would appear as complex conjugated.

In most situations, we do not have information about either spatial distribution or noise statistics of either of the participating modes, and need to extract them from the measurements. This may not be possible under general conditions, since the contributions of all modes can be combined into one simple functional dependence:
\begin{align}
	\label{eq:measured_variance_simplified_form}
	V_m(\theta) = V^+_m \cos^2(\theta - \theta_{m}) + V^-_m\sin^2(\theta - \theta_{m})
\end{align}
where $V^+_m$ and $V^-_m$ are the maximum and minimum quadratures detected
for the $m_\mathrm{th}$ mask respectively and $\theta_m$ is some global mask dependent phase
shift. While these parameters are relatively simple to extract from experimental data (see Fig.~\ref{fig:sqz_detection}), the system of measurements is under-constrained, and we generally do not have enough information to independently extract
$V_k^{\pm}$, ${\cal O}_{km}$, $\theta_k$, and $\theta_{km}$.

Nevertheless, below we consider several important cases, for which we can obtain the quantum mode profiles.

\subsection{Reconstruction of a spatial mode for a single-mode squeezed vacuum}

Let's assume that the input state consists of a squeezed vacuum field in a single unknown spatial mode. 
In this case, Eq.~\ref{eq:detected_variance_multimode} simplifies to
\begin{align}
	V_m(\theta) = 1 + |{\cal O}_{sq_m}|^2 \left( V^+ \cos^2(\theta - \theta_{m}) 
	+ V^- \sin^2(\theta - \theta_{m}) -1 \right)
\end{align}
here we dropped the mode index $k=1$. It is easy to see that minimum and maximum values of the measured quadrature variance are equal to $V_m^\pm =|{\cal O}_{sq_m}|^2 (V^\pm-1)+1$, and therefore we can extract the value of overlap parameter as
\begin{align}
	\label{eq:sq_mode_o}
	{\cal O}_{sq_m} & \propto e^{i \theta_{m}} \sqrt{ {V^+_m -V^-_m} } 
\end{align}
where we omitted the factor $1/\sqrt{V^+ - V^-}$ since it is a common normalization factor for any mask $H_m$.

We can use well established single pixel camera methods for the intensity\cite{Gibson2020} or field\cite{Cuozzo2022spi} spatial distribution reconstruction modified to recover the squeezed field multiplied by the LO field profile, which we call the shaped squeezed field:
\begin{align}\label{eq:shapedsqzfield}
	{\cal U}_{\rm sq}(x,y) = u_{LO}^*(x,y) u_{\rm sq}(x,y)
\end{align}
which is {the main interest of this manuscript}. We can see that projection of the shaped squeezed field to a mask is given by
\begin{align}
	{\cal O}_{sq_m} = \sum_p H_{m}(p) {\cal U}_{\rm sq}(p)\,,
\end{align}
where ${\cal O}_{sq_m}$ is the weight of the $m$th mask in the reconstruction of the shaped squeezed field. 
The above equation can be written in matrix notation as
\begin{align}
	\label{eq:mask_overlap_pixel}
	\vec{{\cal O}_{\rm sq}}^T= \mathbf{H} {\vec{{\cal U}_{\rm sq}}}^{T}\,,
\end{align}
where we move from the continuous two dimensional $xy$ representation
(Eq.~\ref{eq:mask_overlap_integral}) to pixel basis ($p$) and
unfold 2D space to a single column tracking pixel location. To have fully define system, we need
as many independent mask measurements as there are sampled pixels.  The rest is just linear algebra.

The shaped squeezed field can be calculated based on measurements as
\begin{align}
	\label{eq:grand_overlap_inv_matrix}
	{\vec{{\cal U}_{\rm sq}}}^T= \mathbf{H}^{-1} \vec{{{\cal O}_{\rm sq}}}^{T}
\end{align}
Here rows of matrix $\mathbf{H}$ consist of the pixel representations
of the masks. If the $\mathbf{H}^T = \mathbf{H}^{-1}$, such is in the case of the
Hadamard masks, the above equation simplifies to 
\begin{align}
	\label{eq:mask_overlap_hadamard}
	{\vec{{\cal U}_{\rm sq}}}^T= \mathbf{H}^{T} \vec{{{\cal O}_{\rm sq}}}^{T} = 
	\sum_m H^T_{m} {\cal O}_{sq_m}
\end{align}
One potential obstacle comes from the requirement of mask overlaps to have a $\,\pm1$ factor (Eq.~\ref{eq:sq_mode_o} ). This ambiguity is resolved by measuring a complementary mask shape $1-H_m$ that defines an overlap with the unity mask as the reference (${\cal O}_r$). A mask overlap added to its complementary needs to be equal to the reference overlap for any mask, thus constraining the sign. Then a simple comparison of possible permutations of $\pm1$ multipliers for the mask and its complementary one provides the correct sign.

Overall, we have the method to obtain the shaped squeezing field
$u_{LO}^*(x,y) u_{\rm sq}(x,y)$ up to some normalization numerical factor
for the single squeezed mode state.

\subsection{Mode decomposition reconstruction for thermal and squeezed vacuum modes} \label{sec:TSVrec}

Now we consider the input state as combination of one squeezed mode ($\rm sq$
subindex)  and one thermal mode  ($\rm th$ subindex).
In this case we can use Eq.~\ref{eq:noiseNmodes2} to calculate the expected quadrature variance:
\begin{align}
\label{eq:expected_quadrature_variance}
	V_m(\theta) &= 1 + |{\cal O}_{th_m}|^2 (V_{th} -1)+\\&+|{\cal O}_{sq_m}|^2 \left( V^+ \cos^2(\theta -
	\theta_{m}) + V^- \sin^2(\theta - \theta_{m}) -1 \right) \nonumber 
\end{align}

Note that the variance of the thermal state ($V_{th}$) does not depend
on the quadrature angle, and thus its contribution is phase-independent.
This equation obeys general form,
Eq.~\ref{eq:measured_variance_simplified_form}.
Thus we can easily detect $V_m^\pm$ and $\theta_m$ however there is not
enough information to find ${\cal O}_{sq_m}$, ${\cal O}_{th_m}$, $V^\pm$, and
$V_{th}$ from just 3 observables. 
The large thermal mode shifts the observed quantum noise up and dominate it. But Eq.~\ref{eq:expected_quadrature_variance} shows that to obtain squeezed mode overlap we need to track noise contrast (difference between maximum and minimum noise) as shown in Eq.~\ref{eq:sq_mode_o}. This is correct even in the presence of a strong thermal mode.
To reconstruct the shaped squeezed mode ${\cal U}_{sq}$, we can use exactly the same formalism as we used for the case of single squeezed mode above.

Moreover, we assume that thermal mode is much noisier than shot noise, i.e. $V_{th} \gg 1$, and consequently thermal mode variance is much large than squeezed quadrature variance, i.e $V_{th} \gg V^-$. With this assumption 
\begin{align}
	\label{eq:th_mode_o}
	|{\cal O}_{th_m}|^2 & \approx V^-_m 
\end{align}
where we again neglected the common normalization factor $1/V_{th}$.
We can reconstruct intensity overlap of the local
oscillator with the thermal mode, i.e. the shaped thermal field intensity
\begin{align}
    \label{eq:thermal_intensity_overlap}
	| u^*_{LO}(p) u_{th}(p) |^2 = 
	\sum_m H_{m}(p) |{\cal O}_{th_m}|^2
\end{align}
Here we use the fact that variance of the thermal mode is proportional to its intensity and this relationship does not depend on the loss of the system. This allows us to generalize single pixel detector intensity formalism~\cite{Gibson2020,pratik2022}.

\section{Experimental realization}

\begin{figure}
    \centering
    \includegraphics[width=0.48\textwidth]
    {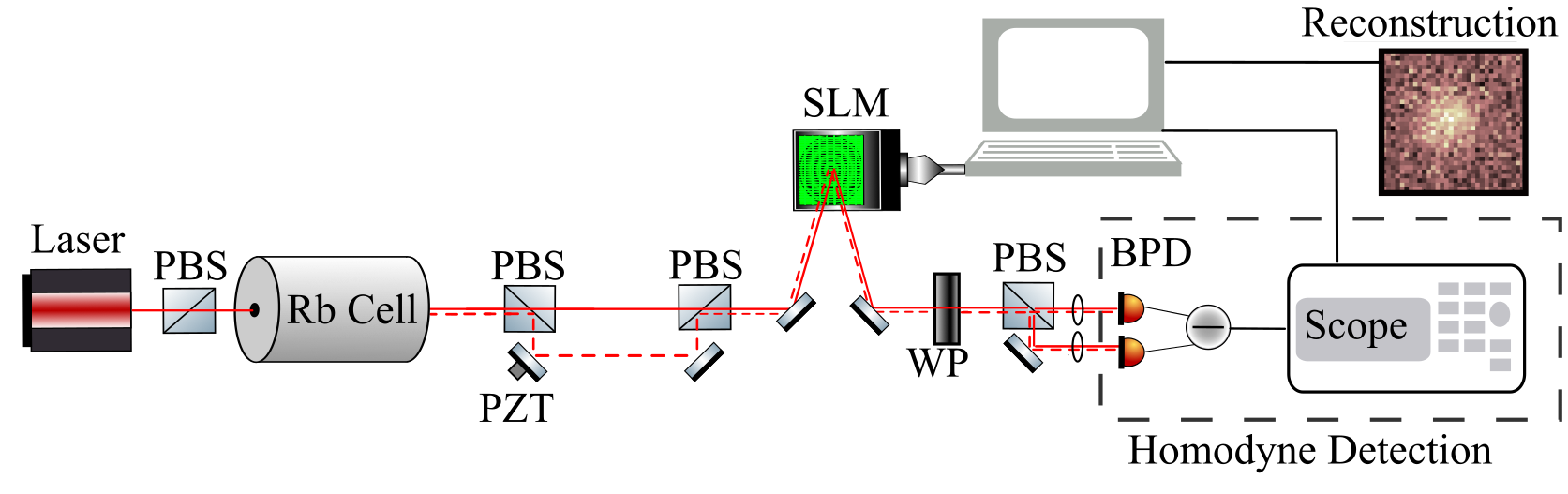}
    \caption{
    Diagram of the experimental setup. See text for the abbreviations.
    }
    \label{fig:ex_setup}
\end{figure}

The experimental apparatus used to illustrate our method is depicted in Fig.~\ref{fig:ex_setup}.
The pump laser beam input power is 7.3~mW at the entrance of the Rb cell and has radius  of $60~\mu$m in the focus (at the center of the cell).
We use a strong linearly polarized pump tuned to the $5S_{1/2} F=2\rightarrow 5P_{1/2}$ transition of the $^{87}$Rb atoms to generate squeezed vacuum field in the orthogonal polarization via polarization self-rotation (PSR) effect~\cite{Cuozzo2022camera,mizhang2016pra}. The output squeezed vacuum field is the input state to the quantum mode spatial profiler, as previous research indicated that this field may contain several squeezed or thermal modes~\cite{mikhailov2009jmo,mizhang2016pra}. For the measurements we reuse the pump field as a LO for the homodyning balanced photodiode detector (BPD) which measures quadrature fluctuations in the squeezed field. We use an interferometer consisting of two polarizing beam splitters (PBS) and two mirrors (one of which is mounted on a PZT transducer) to introduce the controllable
phase shift ($\theta$) between the LO and squeezed field.

We use a phase-only liquid crystal spatial light modulator (SLM, model Meadowlark Optics PDM512-0785). We take advantage of the polarization dependence of the SLM to impose spatial masks only on the squeezed field, without affecting the local oscillator. This arrangement is crucial to reduce the effect of the temporal common phase flicker due to the liquid crystal driving circuit. Since both optical fields propagate and bounce off the SLM together, they see the SLM phase flicker as a common phase which cancels out in the measurement. To introduce a field amplitude mask, we apply a blazing diffraction grating pattern with different modulation depth~\cite{Boyd2013ol_slm_profiling,Cuozzo2022spi} and select its zeroth order. This way we can controllably apply ``on'' or ``off'' patterns of the Hadamard mask basis set to shape the squeezed field. Technically, we need masks with 1 and -1 amplitudes for the Hadamard patterns. As -1 intensities are physically not feasible, we use 1 and 0 patterns and their complementary, following a well established technique for single pixel camera detectors~\cite{Gibson2020}. After the SLM, the unchanged LO and masked squeezing field enter the homodyning BPD, and we record the squeezed field quadrature variance (noise level) with a spectrum analyzer.

We measure noise level as a function of the LO phase for every mask
($V_m(\theta)$), see Fig.~\ref{fig:sqz_detection}, and extract maximum noise levels $V_m^+$, minimum noise levels $V_m^-$, and the
corresponding phase shift $\theta_m$ for every mask, 
as shown in Eq. \ref{eq:measured_variance_simplified_form}.
A blank mask with no modifications to the input squeezed beam is used to define a reference phase with the LO.
From this measurement using
Eqs.~\ref{eq:sq_mode_o}   and \ref{eq:th_mode_o}, we are able to reconstruct the mask overlap for squeezed (${\cal O}_{sq_m}$) and thermal (${\cal O}_{th_m}$) fields.
Once we know this, we reconstruct the
shaped squeezed and thermal fields using Eqs.~\ref{eq:mask_overlap_hadamard} and~\ref{eq:thermal_intensity_overlap}.

\section{Experimental Mode Reconstructions}

\begin{figure}
    \centering
    \includegraphics[width=0.48\textwidth]
    {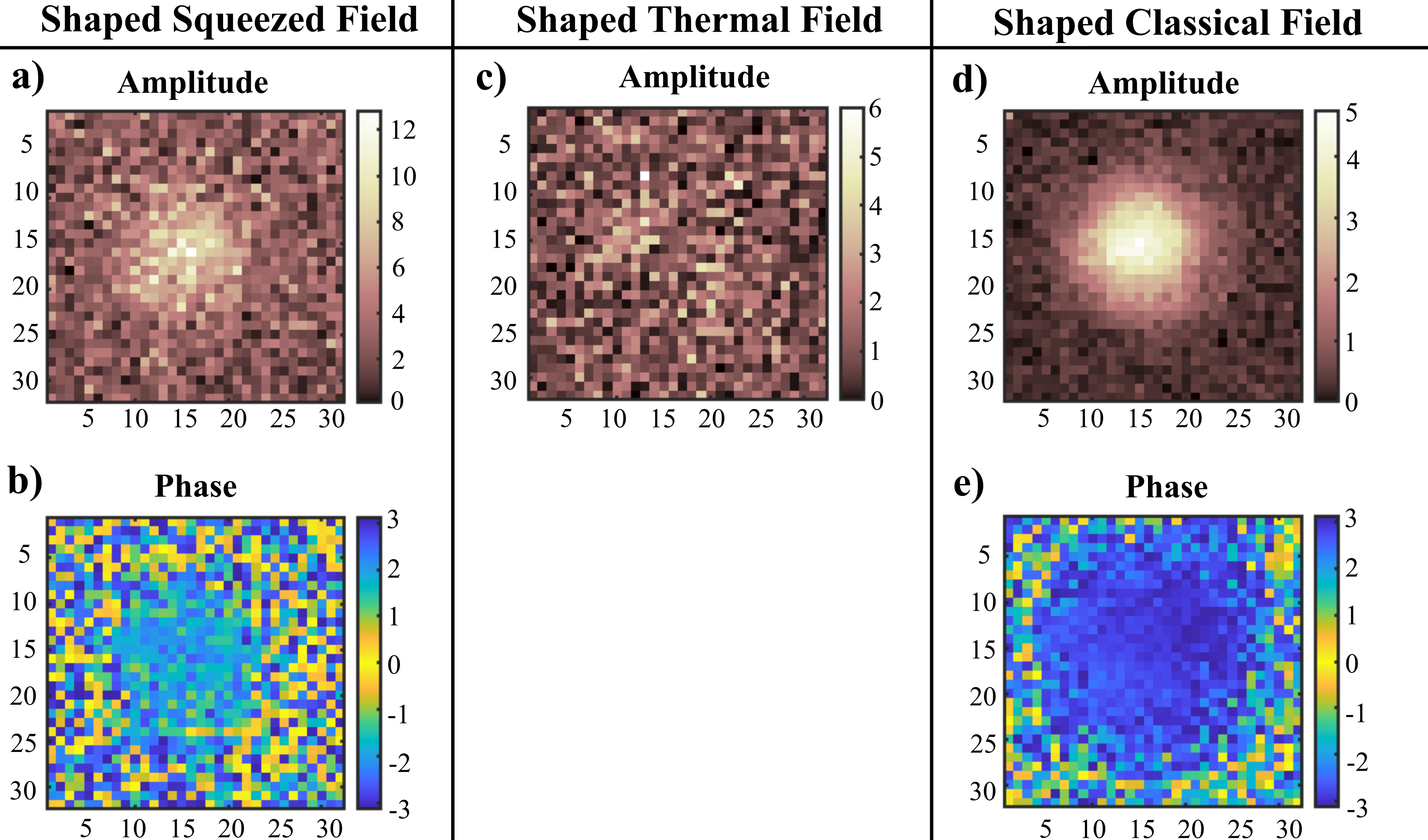}
    \caption{
    A low temperature reconstruction (65ºC) where a) is the amplitude of the shaped squeezed field b) is the squeezed phase c) is the amplitude of the shaped thermal field and d) and e) are the amplitude of the shaped classical field and phase reconstructions respectively.
    Classical field images (recovered with methods described in~\cite{Cuozzo2022spi}) are included to provide comparison.
Phase colorbars are in radians.
Quantum fields amplitude colorbars are proportional to the square root of quantum noise variance.
}
    \label{fig:65recon}
\end{figure}

\begin{figure}
    \centering
    \includegraphics[width=0.48\textwidth]
    {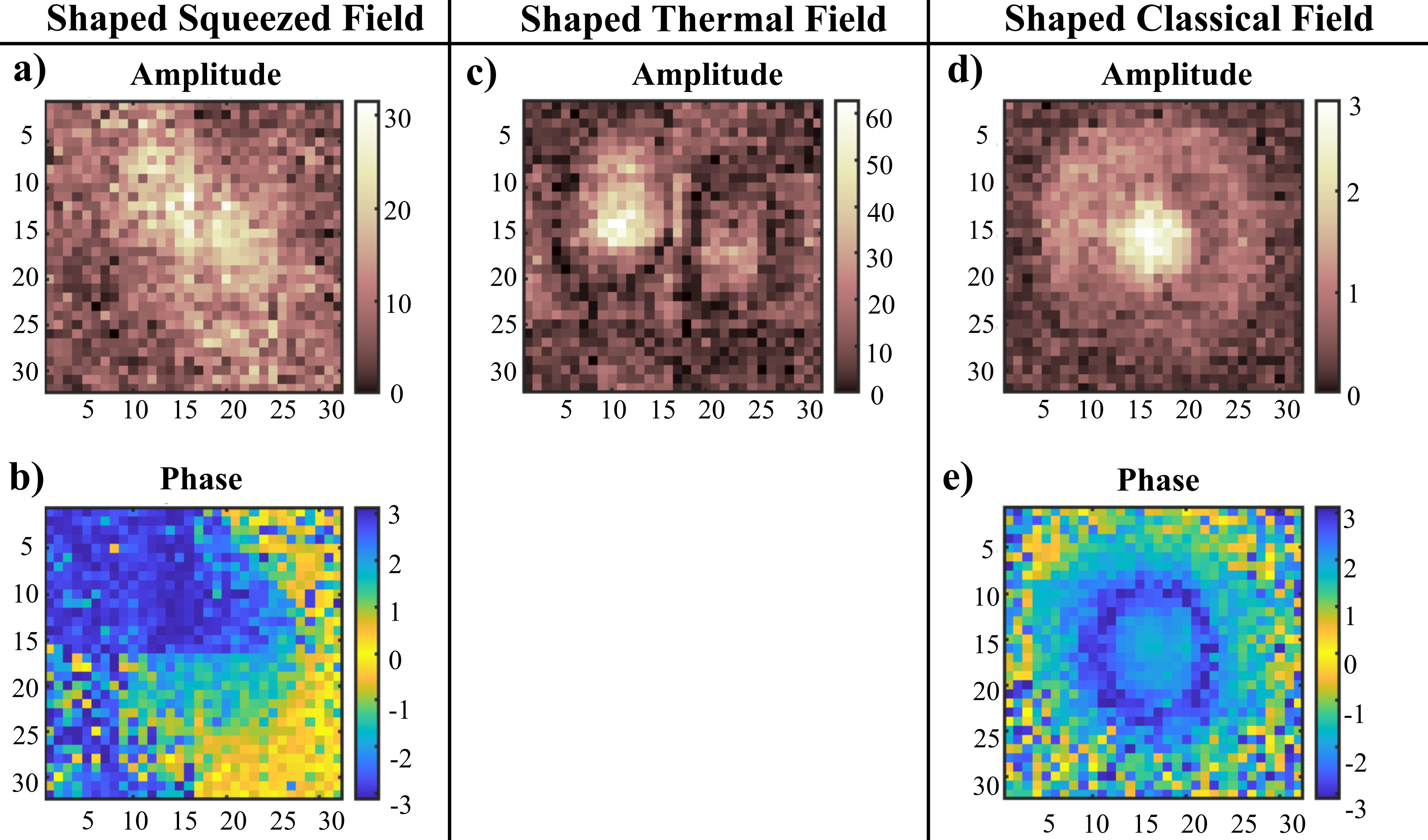}
    \caption{
    A high temperature reconstruction (80ºC) where a) is the amplitude of the shaped squeezed field b) is the squeezed phase c) is the amplitude of the shaped thermal field and d) and e) are the amplitude of the shaped classical field and phase reconstructions respectively.
    Classical field images (recovered with methods described in~\cite{Cuozzo2022spi}) are included to provide comparison.
Note the thermal shape difference (c) compared to Fig. \ref{fig:65recon}c. Phase colorbars are in radians. 
Quantum fields amplitude colorbars are proportional to the square root of quantum noise variance.
}
    \label{fig:80recon}
\end{figure}

PSR squeezing makes a potent subject for the mode decomposition analysis, as many previous experiments demonstrated that it is far from pure, and it is plagued by excess noise~\cite{hsu_effect_2006,mikhailov2009jmo,mizhang2016pra} that increases with temperature of Rb vapor. The spatial mode analysis can shine light on the nature of the excess noise. In particular, we assume that the optical field coming out of the Rb cell consists of a single-mode squeezed vacuum and some thermal noise mode. Previous measurements suggest that shapes of these two modes do not match each other. To distinguish between them we run the mode decomposition analysis for two different Rb cell temperatures: $T=65^o$C, for which the maximum PSR squeezing is detected, and we suspect relatively small contribution from the thermal noise
as this low temperature regime is close to the single squeezed mode~\cite{zhangPRA2017,nicktheorypaper},
and at $T=80^o$C, for which the excess noise dominates
due to the significant addition of the thermal mode. For a direct comparison, see Fig.~\ref{fig:65recon}a,c where squeezing reconstruction has larger noise values and Fig~\ref{fig:80recon}a,c where thermal reconstruction has larger noise values compared to the squeezed amplitude.


 Fig.~\ref{fig:65recon} and Fig.~\ref{fig:80recon} present the 32x32 pixel reconstructions of the squeezed vacuum output that follows the analysis described in Sec.\ref{sec:TSVrec}. Each figure has three distinct columns. The first column shows the amplitude and phase of the overlap between the squeezed mode and the LO, reconstructed using Eq.~\ref{eq:sq_mode_o}. The second column shows the thermal mode shaped intensity reconstructed with Eq.~\ref{eq:thermal_intensity_overlap}.
Note the thermal state by definition has  no phase dependence. This is used as an implicit assumption during reconstruction.
Finally, the last column shows the classical reconstruction~\cite{Cuozzo2022spi} using
 a small leakage of the classical LO field into the squeezing polarization due to  the limited extinction ratio of the polarizing beam displacer).

The lower temperature corresponds to a lower atomic density and weaker nonlinear effect
which is in charge of squeezing and output mode structure\cite{nicktheorypaper}.
The reconstruction at ${65^\circ}$C temperature (Fig.~\ref{fig:65recon}) shows a clear fundamental Gaussian beam shape in both classical (Fig.~\ref{fig:65recon}d,e) and quantum (Fig.\ref{fig:65recon}a,b) reconstructions. This is expected, since the squeezing is generated in the  mode very similar to the LO which was used as a pump for the squeezer~\cite{mikhailov2012jmo_sq_filter,HorromJPB12,mizhang2016pra,nicktheorypaper, zhangPRA2017}.
At ${65^\circ}$C we observe -2.0~dB of squeezing (noise suppression relative to the shot noise level) directly out of the Rb cell. Due to some absorption in optical  elements such as polarizers and less than 100\% reflection off the SLM, this amount of squeezing is reduced to -0.5~dB when the squeezing propagates through the imaging optics (see Fig.~\ref{fig:ex_setup}). We also detect about 5.7~dB of antisqueezing at the detector after passing through the imaging optics, hinting about the thermal noise presence. While we cannot predict the shape of the thermal mode, we must assume that it occupies similar space as the squeezed vacuum as we observe its negative effect on observed squeezing noise~\cite{mizhang2016pra,nicktheorypaper}. This prediction is supported by the measured thermal mode profiles.

To increase atomic density we raise the Rb cell temperature to ${80^\circ}$C. At this high temperature, we no longer have any squeezing (measured $V_m^-$ exceeds the shot noise level) as the minimum noise  is 2.7~dB above shot noise (due to increased contribution of the thermal mode) and the maximum noise is 11.5~dB above the shot noise after passing through the imaging optics. This noise increase is expected with higher temperatures.

When compared to the low temperature reconstruction Fig.~\ref{fig:65recon}, we see a spatial mode change in both classical and quantum reconstructions (see Fig.~\ref{fig:80recon}). In the classical fields overlap reconstruction, an additional ``ring'' appears (Fig.\ref{fig:80recon}e), likely due to self-defocusing of the laser field in hot atomic vapor. The quantum reconstructions (Fig.\ref{fig:80recon}a,b) also show modification of the original Gaussian, even though they suffer from some digital ``boxiness'' that is highly dependent on post-processing phase choices.  However, even the imperfectly reconstructed thermal mode shape (Fig.\ref{fig:80recon}c) (that is phase-independent) is very distinct from the classical shapes, as  two ``lobes'' appear. One can notice similar two-lobe structure even in the low-temperature thermal mode reconstruction (Fig.\ref{fig:65recon}c), albeit much less obvious.

The magnitude of the reconstructed fields is proportional to input squeezing and thermal variances (recall that we did not normalize by $\sqrt{V^+-V^-}$ and $V_{\rm th}$ in Eqs.~\ref{eq:sq_mode_o} and \ref{eq:th_mode_o}). Thus we can see that at higher atomic densities
a noisier (higher input variance) field is generated.


We would like to note that it is possible to get higher resolution
images, since we were mainly limited by the acquisition time for each mask and speed of SLM (Meadowlark PDM512) liquid crystal settling, which was the bottleneck of our setup. It takes about 45 minutes to collect a 32x32 pixel reconstruction.

\section{Conclusion}
We demonstrated a method to reconstruct spatial profile of an optical field consisting of several quantum noise modes with different transverse profiles. The proposed formalism is general but we specifically considered the case of a single-mode  squeezed vacuum field, alone or with some contribution of a thermal mode. We applied this analysis for the squeezed vacuum generated in Rb vapor due to PSR effect, and observe signs of thermal noise emergence at higher temperatures, as expected from previous experimental results.
Potentially, when measurements extract enough information about the covariance matrix a back transformation can be applied and the initial covariance matrix can be exactly reconstructed. We can verify the reconstruction fidelity when the process finds a diagonal covariance matrix. 
The developed profiler technique has potential  use in many quantum communication and precision measurement applications, where exact mode matching with an unknown quantum mode is necessary for high-fidelity quantum state detection. 

\section*{Author declarations}\label{DECLARATIONS}

\subsection*{Conflict of Interest}
The authors have no conflicts to disclose.

\section*{Data availability}\label{DATA_AVAILABILITY}
The experimental data that support the findings of this study are available from the corresponding author upon reasonable request.

\section*{Funding}\label{sec:funding}
Air Force Office of Scientific Research (FA9550-19-1-0066).
 
\section*{References}\label{sec:refernces}


\begin{thebibliography}{43}
\bibitem [{sym()}]{symplectictransform}%
  \BibitemOpen
  \href@noop {} {}\bibinfo {note} {Action of any Gaussian process can be
  represented by a symplectic matrix $S$ with the transformations $ \bar{x}
  \longrightarrow S\bar{x}, V \longrightarrow S V S^T $. These transformations
  connect the moments of the states before and after the process, and since the
  moments completely describe the state these transformations completely
  describe the state after the process.}\BibitemShut {Stop}%
\end{thebibliography}
\end{document}